\documentclass[prl,aps,reprint,showpacs,raggedbottom,nofootinbib,superscriptaddress,floatfix]{revtex4-1}
\usepackage{bm,dcolumn}
\usepackage{amsmath,amsfonts,amssymb,amsthm,amscd} 
\usepackage{graphicx}
\usepackage{verbatim}
\usepackage[section]{placeins}
\usepackage{color}
\usepackage{soul}
\usepackage{xparse}
\usepackage{physics}
\usepackage{siunitx}

\NewDocumentCommand\ZZ{+u{ZZ}}{\ignorespaces}

\begin{document}
\sloppy
\title{Spontaneous avalanche dephasing in large Rydberg ensembles}

\author{T. Boulier}
\affiliation{Joint Quantum Institute, National Institute of Standards and Technology and the University of Maryland, College Park, Maryland 20742 USA}
\affiliation{Laboratoire Charles Fabry, Institut d’Optique Graduate School, CNRS, Universit\'e Paris-Saclay, 91127 Palaiseau cedex, France}

\author{E. Magnan}
\affiliation{Joint Quantum Institute, National Institute of Standards and Technology and the University of Maryland, College Park, Maryland 20742 USA}
\affiliation{Laboratoire Charles Fabry, Institut d’Optique Graduate School, CNRS, Universit\'e Paris-Saclay, 91127 Palaiseau cedex, France}

\author{C. Bracamontes}
\affiliation{Joint Quantum Institute, National Institute of Standards and Technology and the University of Maryland, College Park, Maryland 20742 USA}

\author{J. Maslek}
\affiliation{Joint Quantum Institute, National Institute of Standards and Technology and the University of Maryland, College Park, Maryland 20742 USA}

\author{E. A. Goldschmidt}
\affiliation{United States Army Research Laboratory, Adelphi, Maryland 20783 USA}

\author{J. T. Young}
\affiliation{Joint Quantum Institute, National Institute of Standards and Technology and the University of Maryland, College Park, Maryland 20742 USA}

\author{A. V. Gorshkov}
\affiliation{Joint Quantum Institute, National Institute of Standards and Technology and the University of Maryland, College Park, Maryland 20742 USA}
\affiliation{Joint Center for Quantum Information and Computer Science, National Institute of Standards and Technology and the University of Maryland, College Park, Maryland 20742 USA}

\author{S. L. Rolston}
\affiliation{Joint Quantum Institute, National Institute of Standards and Technology and the University of Maryland, College Park, Maryland 20742 USA}

\author{J. V. Porto}
\email{porto@umd.edu}
\affiliation{Joint Quantum Institute, National Institute of Standards and Technology and the University of Maryland, College Park, Maryland 20742 USA}

\date{\today}

\begin{abstract}
Strong dipole-exchange interactions due to spontaneously produced contaminant states can trigger rapid dephasing in many-body Rydberg ensembles [E. Goldschmidt \textit{et al.}, PRL 116, 113001 (2016)]. Such broadening has serious implications for many proposals to coherently use Rydberg interactions, particularly Rydberg dressing proposals. The dephasing arises as a runaway process where the production of the first contaminant atoms facilitates the creation of more contaminant atoms. Here we study the time dependence of this process with stroboscopic approaches. Using a pump-probe technique, we create an excess ``pump" Rydberg population and probe its effect with a different ``probe" Rydberg transition. We observe a reduced resonant pumping rate and an enhancement of the excitation on both sides of the transition as atoms are added to the pump state. We also observe a timescale for population growth significantly shorter than predicted by homogeneous mean-field models, as expected from a clustered growth mechanism where high-order correlations dominate the dynamics. These results support earlier works and confirm that the time scale for the onset of dephasing is reduced by a factor which scales as the inverse of the atom number. In addition, we discuss several approaches to minimize these effects of spontaneous broadening, including stroboscopic techniques and operating at cryogenic temperatures. It is challenging to avoid the unwanted broadening effects, but under some conditions they can be mitigated.
\end{abstract}

\maketitle

\section{Introduction}
The dense level spacing of Rydberg atoms can be both a blessing and a curse. The strong polarizability of these highly excited states gives rise to giant interactions which, along with extended lifetimes, make them appealing candidates for engineering a variety of many-body Hamiltonians. Strong interactions are central to many-body experiments, including quantum magnetism~\cite{barredo2015,labuhn2016tunable} \ZZ, quantum thermalization~\cite{Lee2013}, as well as ZZ and quantum information processing~\cite{weimer2010,Pohl2011, Saffman2010}. Numerous proposals with varying levels of controllability and timescales use Rydberg interactions~\cite{heidemann07a,barredo2015,bendkowsky2009,urban2009,pupillo2010,Glaetzle2012,Bouchoule2002,Glaetzle2014,vanBijnen2015,Lee2013a}. Importantly, Rydberg dressing proposals aim to engineer long-lived, interacting states by weakly admixing a Rydberg state with the ground state using off-resonant laser coupling~\cite{Johnson2010,henkel10a,pupillo2010,honer10a,Glaetzle2012,vanBijnen2015,Glaetzle2015,Bouchoule2002,Lee2013a,Dauphin2012,Glaetzle2014}. While Rydberg dressing has been demonstrated for small atomic ensembles (100 atoms or fewer) and for short timescales~\cite{Bloch, jau2016entangling}, in the many-body regime success at coherent off-resonant dressing has been more elusive~\citep{Killian16,aman2016trap,AB,pohl,Balewski2014}.

A major limitation not addressed by these proposals is a blackbody-induced avalanche dephasing effect, where the runaway production of atoms in strongly interacting contaminant states of opposite parity significantly decreases the available coherence time~\cite{aman2016trap,AB,Bloch,Killian16}. Given the potential applications of coherent Rydberg control, it is crucial to take such dissipative dynamics into account. Therefore, charting the properties of this source of decoherence is important for the development of Rydberg-based quantum computation and simulation. This work is such an effort, focusing on the dephasing's time-dependence and impact on current research.

In order to study the time dependence of this broadening mechanism, we use a pump-probe technique where excitation to a ``pump" Rydberg state induces broadening on a separate, ``probe" Rydberg transition. Additionally, we study the early time dynamics on a single Rydberg transition using both a stroboscopic approach and photon-counting measurements and show that homogeneous mean-field assumptions underestimate the speed of the avalanche process by an order of magnitude. This is a strong indication of the highly-correlated, locally seeded nature of this avalanche process~\cite{ates2007many,jeremy}, which is relevant to the question of bistability in large, strongly-interacting systems~\cite{letscher2016bistability}. Both this correlated turn-on and the observation of a giant steady-state broadening rely on anti-blockade mechanisms~\cite{letscher2017anomalous,letscher2016bistability}. Finally, we discuss experimental parameters that might be tuned to extend coherence times, including operation at cryogenic temperatures and stroboscopic application of the excitation light.

\section{Experiment}
In a room-temperature radiation environment, a Rydberg atom has a significant probability to undergo a blackbody-stimulated transition to a Rydberg state of opposite parity~\cite{Gallagher98,cooke1980effects}, most probably changing the principal quantum number by 0 or 1. A single atom in such a state has extremely strong dipole exchange interactions with subsequently excited Rydberg atoms of the original parity~\cite{Park2011,Gunter2013,Brekke2012,Gallagher16}. For two atoms separated by $\bm{r}$, the interaction energy has the form~\cite{Afrousheh2004,browaeys2016experimental}
\begin{equation}
\hat{V}_{dd}(\bm{r}) = \frac{1}{4\pi\epsilon_0}\frac{\bm{\hat{d_1}}\cdot\bm{\hat{d_2}}-3(\bm{\hat{d_1}\cdot n})(\bm{\hat{d_2}\cdot n})}{|\bm{r}|^3},
\end{equation}
with $\bm{n} = \bm{r}/|\bm{r}|$, and where $\bm{\hat{d}}_i$ is the electric dipole operator of atom $i$. Because of the strength of the dipole interaction (scaling with the fourth power of the principal quantum number) and the $1/\bm{r}^3$ long-range character, a single perturber atom can affect many other atoms, potentially the entire sample. The contaminant atom shifts the transition energies of the atomic ensemble and modifies the remaining atoms' spectroscopic properties. 

\begin{figure}[!t]
\includegraphics[width=8.6cm]{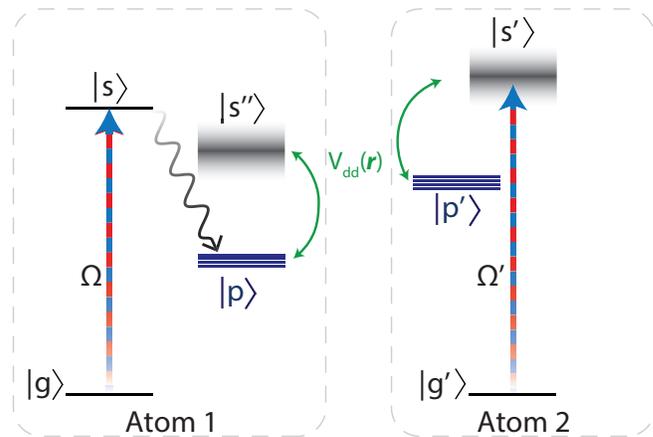}
\caption{\textbf{Interaction mechanism:} Blackbody-dominated decay from the Rydberg state produces nearby p-states ($\ket{p}$), which trigger an off-diagonal dipole-exchange interaction of the form $\ket{p,s'}\rightarrow\ket{s'',p'}$. This creates a dipole-dipole exchange interaction between two different Rydberg s-states. For some orientation and distance \textbf{r} between the dipoles, this brings atom 2 into resonance and the excitation is enhanced (anti-blockade). This energy shift inhomogeneously broadens the transition in the sample, as represented by the gray shading on $\ket{s'}$ and $\ket{s''}$. If $\ket{s}=\ket{s'}$, we are in the ``self-broadening'' situation where a single s-state transition is broadened via the p-states created from it.}
\label{AB-Cartoon}
\end{figure}

Fig.~\ref{AB-Cartoon} illustrates the generic case where p-states originating from the decay of one s-state population dephase a different s-state transition (``cross-broadening''). In the case where there is only one s-state involved ($\ket{s}=\ket{s'}=\ket{s''}$ in Fig.~\ref{AB-Cartoon}), that state is dephased by the product of its own decay (``self-broadening'').

We focus first on the simpler self-broadening mechanism: an off-resonant excitation (central to Rydberg dressing proposals) can resonantly excite atoms in the presence of contaminants, facilitating production of more contaminant Rydberg population~\cite{Urvoy2015} on both sides of the resonance. The process results in a rapid dephasing of the sample~\cite{raitzsch2009investigation}.  The interacting, dynamical excitation process can be complicated~\cite{jeremy,letscher2016bistability}, but since it is triggered by the {\em first} blackbody decay, its onset occurs on a time scale $N$ times faster than the single atom scattering rate would indicate; here $N$ is the number of atoms participating in the Rydberg excitation~\cite{AB,Bloch}. The subsequent runaway excitation quickly reaches steady-state Rydberg populations, effectively broadening the transition. Previous work explored the impact of similar, controlled contaminant Rydberg populations~\cite{Gallagher98,Park2011,Gunter2013,Brekke2012}. However, the systematic appearance of uncontrolled contaminants has only recently been recognized~\cite{Bloch,AB} and a dynamic description was missing.

Since the resonant dipole interaction with the contaminant atoms exists for any of the various Zeeman or hyperfine states, cross-broadening should occur between transitions to two different Rydberg states, as shown in  Fig.~\ref{AB-Cartoon}.  Contaminant population in state $\ket{p}$ produced by blackbody decay from $\ket{s}$ can broaden subsequent excitation to a separate state $\ket{s^\prime}$. This suggests the possibility to study the broadening mechanism using a strong ``pump'' excitation to create contaminant Rydberg population, and a weak ``probe'' transition to measure the population's impact on linewidth.  This approach has the benefit that the amount and time dependence of the contaminant population can be controlled independently from the probing transition~\cite{letscher2017anomalous}. Here we use excitation to two different Rydberg hyperfine states to study the broadening mechanism. 

\subsection{Experimental Details}

\begin{figure}[!t]
\includegraphics[width=8.6cm]{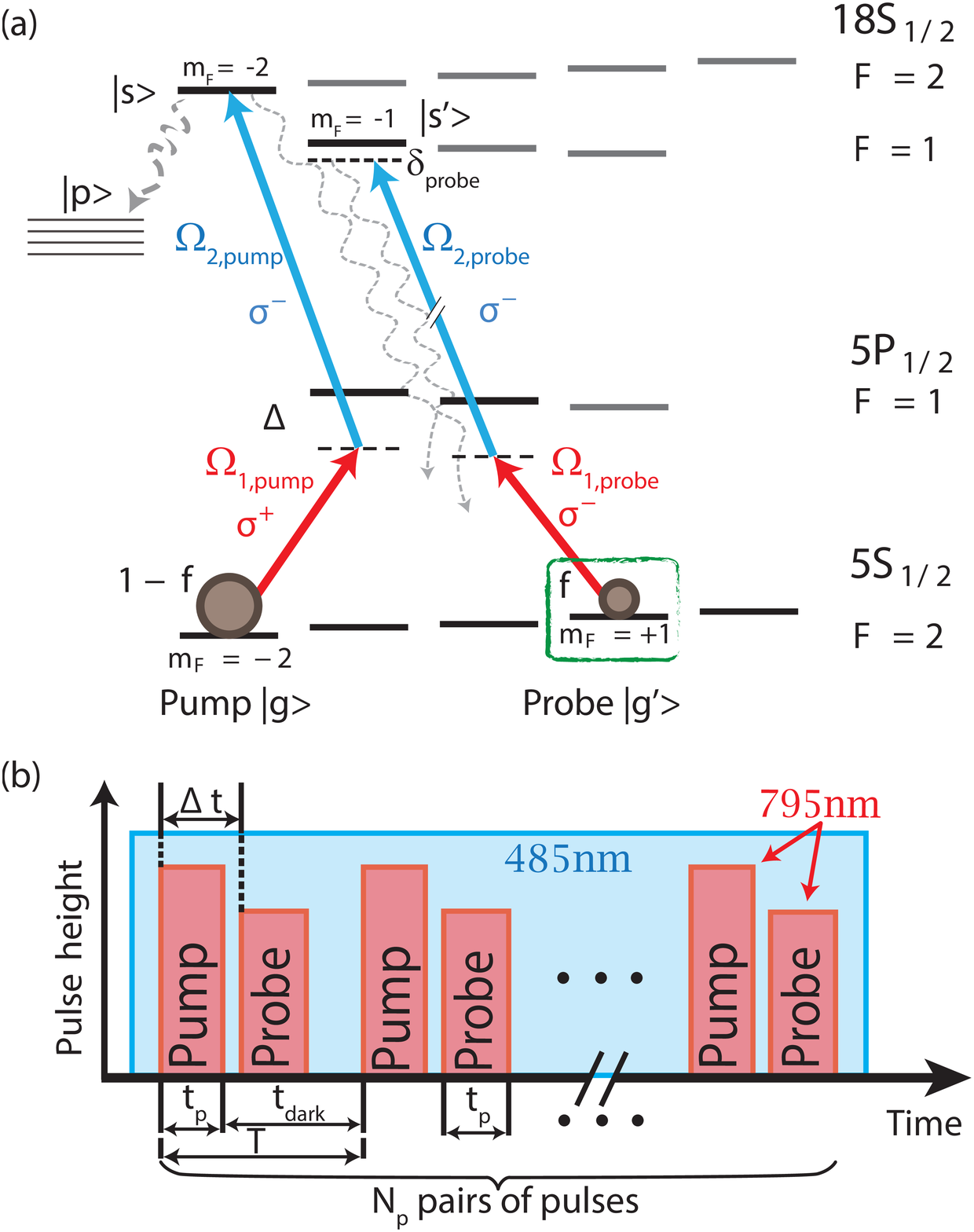}
\caption{\textbf{Excitation scheme. (a)} The ground state population is divided into a fraction $f$ in the ``probe state" $\ket{g^\prime}$, leaving the fraction $1-f$ in the ``pump state" $\ket{g}$. We chose $f=0.25$, small enough for the contaminant p-atoms to come predominantly from the pump, but high enough to have a sufficient signal to noise. The excitation to $18S$ is done via a two-photon transition with the $5P_{1/2}$ intermediate state: the intermediate detuning $\Delta$ is much greater than the single-photon Rabi frequencies and than the two-photon detuning $\delta_\textrm{probe}$ (not to scale on this figure). The pump population is driven resonantly to $\ket{s}=\ket{18S,2,-2}$ with two-photon Rabi frequency $\Omega_\textrm{pump}$. The two-photon detuning $\delta_\textrm{probe}$ of the probe transition to $\ket{s^\prime}$ is scanned to record depletion spectra in the ground probe state (highlighted by a green square). \textbf{(b)} To observe effects of cross-broadening, we chop both $\si{795~\nano\meter}$ beams (probe and pump) into pulse trains, which can be delayed relative to one another. The $\si{485~\nano\meter}$ light is kept on for the whole pulsing sequence.}
\label{Schematics}
\end{figure}

The details of our experimental setup are described in~\cite{AB} and~\cite{BEC}. Briefly, the apparatus is used to create $^{87}$Rb Bose-Einstein condensates (BEC) of $N \approx 4\times 10^4$ atoms initially in the $\ket{F=1, m_F=-1}$ ground state. Arbitrary fractions of the atoms can be transferred to any other state within the ground hyperfine manifold via microwave rapid adiabatic passage. We quantize the inter-atomic distances by loading the $^{87}$Rb BEC into a 3D optical lattice made with $812~\si{\nano\meter}$ light, resulting in a lattice spacing of $\si{406~\nano\meter}$~\cite{BEC}. We then excite atoms to the $18S_{1/2}$ state using a two-photon transition via the $5P_{1/2}$ intermediate state (Fig.~\ref{Schematics}). The van der Waals (vdW) blockade radius is defined as the distance below which the interactions are larger than the collective two-photon Rabi frequency~\cite{saffman}: for the present experimental scheme it is smaller than the lattice spacing. This corresponds to a vdW energy $E_\textrm{vdW}=2\pi\times 2.1~\si{\mega\hertz}$. We note that the lattice is also instrumental in suppressing supperadiant Rayleigh scattering~\cite{Inouye1999}. The $\si{485~\nano\meter}$ light coupling $5P$ to $18S$ is common to both (pump and probe) two-photon transitions. The intermediate state detuning is $|\Delta/2\pi|\approx \si{240~\mega\hertz}$ and the single-photon Rabi frequencies are independently calibrated: $\Omega_1/2\pi=0$~MHz to $10$~MHz (on the $5S-5P$ transition) and $\Omega_2/2\pi\approx25$~MHz (on the $5P-18S$ transition). The two lasers are stabilized to the same high-finesse optical cavity with $<10$~kHz linewidth, and are polarized and tuned to couple the ground hyperfine states to the desired $\ket{18S_{1/2},F_i,m_{F\:i}}$ states (where $i=\{$probe, pump$\}$). Separate pump and probe excitation laser beams at the $5S-5P$ transitions can be applied, while we apply only one $5P-18S$ beam. The two-photon detunings $\delta_i$ and Rabi frequencies $\Omega_i=\Omega_{1,i} \Omega_{2,i}/2\Delta$ are independently tunable for each transition. The post-excitation populations in all the ground hyperfine states (including those optically pumped to states different from $\ket{g}$ and $\ket{g^\prime}$, the pump and probe states) are separated in time-of-flight with a Stern-Gerlach magnetic field gradient and measured via absorption imaging. We scan the various parameters and count the fractional population remaining in the ground states $\ket{g}$ and $\ket{g^\prime}$ to obtain spectra of the transitions. The values and uncertainties of the widths presented in this work are derived from fits of these spectra, each containing at least 50 datapoints.

Fig.~\ref{Schematics}(a) describes the pumping scheme: In the ground state, the atomic sample is divided into a small probe population (fractional density $f$) in a state $\ket{g^\prime}= \ket{5S, F=2, m_F=1}$ within the ground manifold, and a large pump population (fractional density $1-f$) in the state $\ket{g} = \ket{5S, F=2, m_F=-2}$. The atoms are then excited to their respective pump or probe Rydberg state within the $18S$ hyperfine manifold via two-photon excitation: the pump is tuned to the $\ket{g} \rightarrow \ket{s}=\ket{18S, F=2, m_F=-2}$ transition while the probe is tuned near the $\ket{g^\prime} \rightarrow \ket{s^\prime} = \ket{18S, F=1, m_F}$ transition, which is shifted by the $18S$ hyperfine splitting $\Delta_\textrm{HF}=2\pi\times 10$~MHz. Here, $m_F=\{1,0,-1\}$ depends on the choice of probe laser polarization. We can pulse both $\si{795~\nano\meter}$ beams (probe and pump), which can be delayed by $\Delta t$ relative to one another [see Fig.~\ref{Schematics}(b)]. The $\si{485~\nano\meter}$ light is kept on for the entire pulse sequence.

\begin{figure}[!t]
\includegraphics[width=8.6cm]{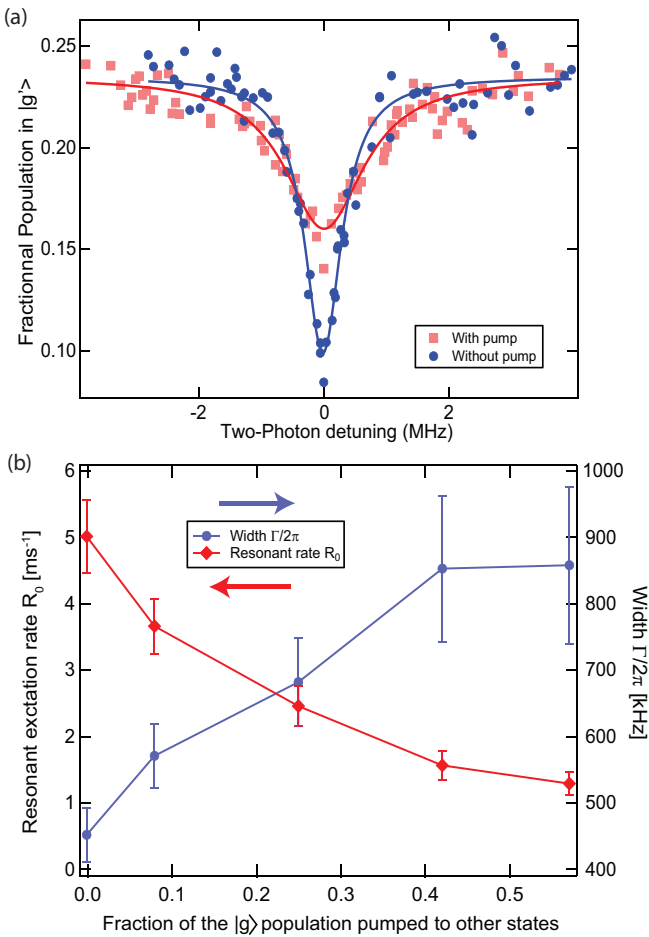}
\caption{\textbf{Cross-state broadening. (a)} Example of broadening due to the presence of pump-induced Rydberg population. The blue (red) curve is the probe spectrum with the pump turned off (on). The fits are Lorentzian, and are used to estimate the resonant rates, the widths, and their uncertainties throughout this work. \textbf{(b)} Probe transition widths and resonant excitation rates as a function of the fraction of $\ket{g}$ that was pumped to other states during the $\si{300~\micro\second}$ excitation. We observe a doubling of the width $\Gamma$ and a significant decrease of the resonant excitation rate $R_0$ between no pump and the maximum pumped population.}
\label{results_nopulse}
\end{figure}

We also have access to the time-resolved, total density of $18S$ atoms, through collection of the fluorescence photons emitted on the $5P_{3/2}-5S_{1/2}$ transition~\cite{AB}. The fluorescence scales with the optical pumping signal and is proportional to the number of $18S$ atoms. It is collected by a lens relay system (NA=0.12) with an interference filter to block the $5S-5P_{1/2}$ excitation light, detected by a single photon avalanche diode, time-tagged with 21-ns resolution, and summed over many repetitions. For technical reasons including the fact that this method cannot discriminate between different states of the $18S$ hyperfine manifold, we use it to gain insight on the dynamics of the dephasing for a single transition (i.e. for the case of self-broadening).

\subsection{Cross-broadening}

We first explore the cross-broadening using the $\ket{g^\prime} \rightarrow \ket{s^\prime} = \ket{18S, 1, -1}$ probe transition, Fig.~\ref{Schematics}(a). With a fixed two-photon probe Rabi frequency $\Omega_\textrm{probe}/2\pi = \si{14~\kilo\hertz}\ll \Gamma_0$ and a single, simultaneous ($\Delta t = 0$) excitation time $t_p=300~\si{\micro\second}$ for both transitions, we take excitation spectra on the probe transition at different pump Rabi frequencies $\Omega_\textrm{pump}$. $\Gamma_0/2 \pi = \si{45~\kilo\hertz}$ is the linewidth of the $18S$ transition, including blackbody radiation. Example spectra are shown in Fig.~\ref{results_nopulse}(a) for two cases, with and without pump excitation. The cross-broadened spectra are reasonably well described by Lorentzian functions, and we fit the data to extract the Lorentzian width $\Gamma$ and the amplitude (along with their uncertainties). The average resonant excitation rate $R_0$ of probe atoms is determined from the fitted amplitudes, given the excitation time and the branching ratios to other states in the ground hyperfine manifold. 

The probe spectra show a significant modification of both the resonant excitation rate and the width as a function of $\Omega_\textrm{pump}$. We note that the range of available $\Omega_\textrm{pump}$ is limited by the constraint that the excitation time $t_p$ must be long enough to give good signal-to-noise on the weakly excited probe transition. On the other hand, $\Omega_\textrm{pump}$ should not be so large as to significantly deplete the pump-state population in $\ket{g}$, in which case the steady-state population in Rydberg state $\ket{s}$ would decay during the probe pulse.  

Fig.~\ref{results_nopulse}(b) shows the probe width $\Gamma$ and excitation rate $R_0$ for different $\Omega_\textrm{pump}/2\pi$ ranging from $\si{0~\kilo\hertz}$ to $\si{20~\kilo\hertz}$, plotted as a function of the fraction of atoms pumped out of the state $\ket{g}$ during the excitation time $t_p$. Note that this is proportional to the number of steady-state contaminant atoms, at least for small fractions. The width observed on the probe transition without pump is $\Gamma/2 \pi \simeq 10 \Gamma_0/2\pi = 450~\si{\kilo\hertz}$, indicating some self-broadening. This is a constant effect for fixed $\Omega_\textrm{probe}$, and is compatible with the steady-state $\Omega$-scaling observed previously~\citep{AB} on a single transition,
\begin{equation}
\Gamma\simeq\Omega\sqrt{\beta \rho_0}, \label{Eq-AB}
\end{equation}
where $\rho_0$ 
is the density of ground state atoms and $\beta$ is an effective interaction volume inherent to the Rydberg state defined as
\begin{equation}
\beta=\sum \left|C_3^\textrm{(nP)}\right| b_\textrm{nP}\Gamma_\textrm{nP}^\textrm{-1},
\end{equation}
where the sum is over the $nP$ states that have substantial dipole interactions with, and branching ratio from, the $18S$ state. Here, $b_\textrm{nP}$ are the branching ratios to the $nP$ states, $\Gamma_\textrm{nP}$ their decay rates and $C_3^\textrm{(nP)}$ are effective dipole interaction strengths (including the root-mean-squared average of the angular dependence of $C_3$~\cite{AB}). When the pump is on, the width of the probe transition is increased by up to a factor of two while the amplitude is reduced by four, showing that the pump creates a significant cross-broadening due to the p-state population. The enhanced off-resonant excitation rate in the presence of the pump is indicative of pump-induced facilitation dynamics~\cite{letscher2017anomalous}.
The observation of broadening between independent populations with independent transitions provides evidence for interaction between atoms and rules out superradiant broadening effects.

\begin{figure}[t]
\includegraphics[width=8.6cm]{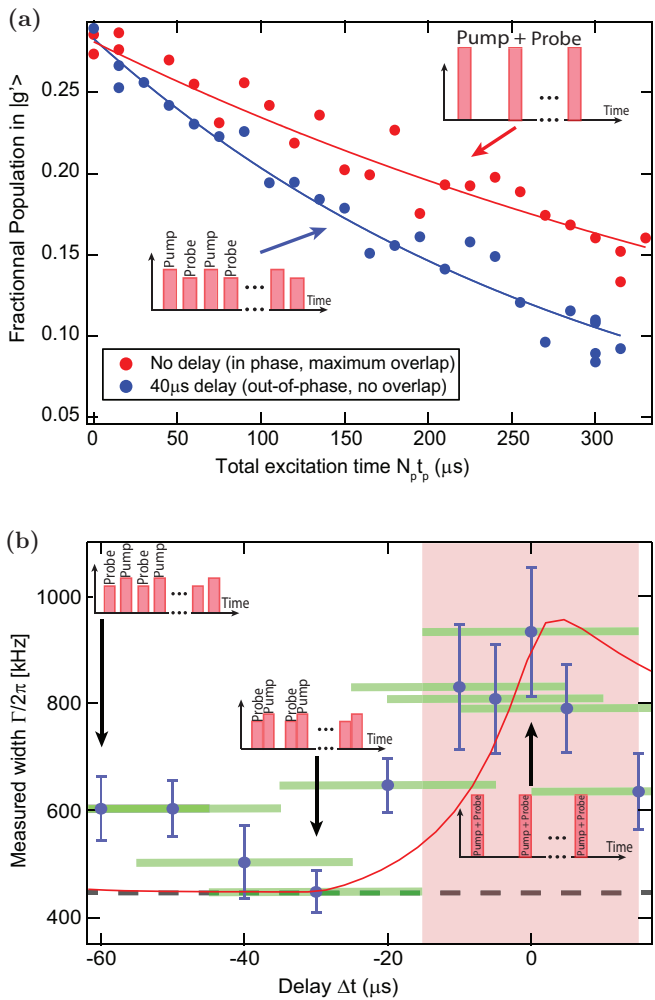}
\caption{\textbf{Delayed pulses experiment. (a)} Pump-induced slowdown of resonant excitation, visible in number of atoms remaining in the probe ground state as a function of excitation time $N_p T$. The two examples here, fitted with exponential decay functions, are for $\Delta t = \si{0~\micro\second}$ (red) and for $\Delta t = \si{40~\micro\second} > t_p$ (blue) and show resonant pumping rates of $R_{\Delta t = \si{0~\micro\second}} = 1.8\pm 0.13~\si{\milli\second^{-1}}$ and $R_{\Delta t = \si{40~\micro\second}} = 3.3\pm 0.14~\si{\milli\second^{-1}}$. \textbf{(b)} Following the pulsing scheme presented in Fig.~\ref{Schematics}(b), we observe a broadening that depends on the overlap of the pulses. The red zones represent the times the pump is on, while the green bands show the probe pulse width. Spectra corresponding to a high p-population are obtained for $\Delta t \simeq \si{0~\micro\second}$ and spectra corresponding to no pump population (pump turned off, marked by the dashed line) are recovered for $|\Delta t| \gtrsim t_p$. This is visible around $\Delta t = \si{-40~\micro\second}$, when the probe is delayed well beyond the lifetime of the Rydberg states. The red line is the result of the nonlinear mean-field theory.}
\label{result_pulsed_delay}
\end{figure}

\subsection{Cross-Broadening Dynamics}
Atoms in s and p states can obviously only interact when present simultaneously. To demonstrate the simultaneous nature of the cross-broadening, we chopped the $795~\si{\nano\meter}$ excitation light for both excitations into pulse trains and studied the effect of non-overlapping pump and probe light. Each pulse had a length of $t_p = 30~\si{\micro\second} \sim 10 \tau_0$ separated by a dark time of $t_\textrm{dark} = 3 t_p$, such that the total period was $T=4 t_p=\si{120~\micro\second}$. For fixed Rabi frequencies $\Omega_\textrm{probe} = 2\pi\times\si{15~\kilo\hertz}$ and $\Omega_\textrm{pump} = 2\pi\times\si{20~\kilo\hertz}$ (corresponding to the largest width seen in Fig.~\ref{results_nopulse}), we vary the delay $\Delta t$ between the probe and the pump pulse trains. We chose $t_p$ to be long compared to the onset of the avalanche dephasing, and the dark time following each pulse in both excitation (probe and pump) to be long with respect to the lifetime ($3-10~\si{\micro\second}$) of the various s and p Rydberg populations.

As expected, the cross-broadening and cross-saturation are absent when the probe and pump pulses are out of phase. Fig.~\ref{result_pulsed_delay}(a) shows the resonant decay of $\ket{g^\prime}$ at fixed $\delta_\textrm{probe}/2\pi=\delta_\textrm{pump}/2\pi=\si{0~\mega\hertz}$ for two delays: $\Delta t=0~\si{\micro\second}$ (overlap) and $\Delta t=40~\si{\micro\second}$ (no overlap). The total excitation time is varied by changing the number of pulses, and the rates are extracted from exponential fits.  The observed rate for simultaneous excitation ($R_{\Delta t = \si{0~\micro\second}} = 1.8\pm 0.13~\si{\milli\second^{-1}}$) is slower than for non-overlapping excitation ($R_{\Delta t = \si{40~\micro\second}} = 3.3\pm 0.14~\si{\milli\second^{-1}}$). Fig.~\ref{result_pulsed_delay}(b) shows the extracted width for probe spectra taken at different $\Delta t$. We observe a variation of the linewidth as a function of the delay. For $\Delta t=0~\si{\micro\second}$ we recover the same result as with the maximum p-population in Fig.~\ref{results_nopulse}(b). We also observe the same width as without any pump light when the two pulse trains are non-overlapping and separated by long enough dark times for their Rydberg populations to decay.

\begin{figure}[t]
\includegraphics[width=8.6cm]{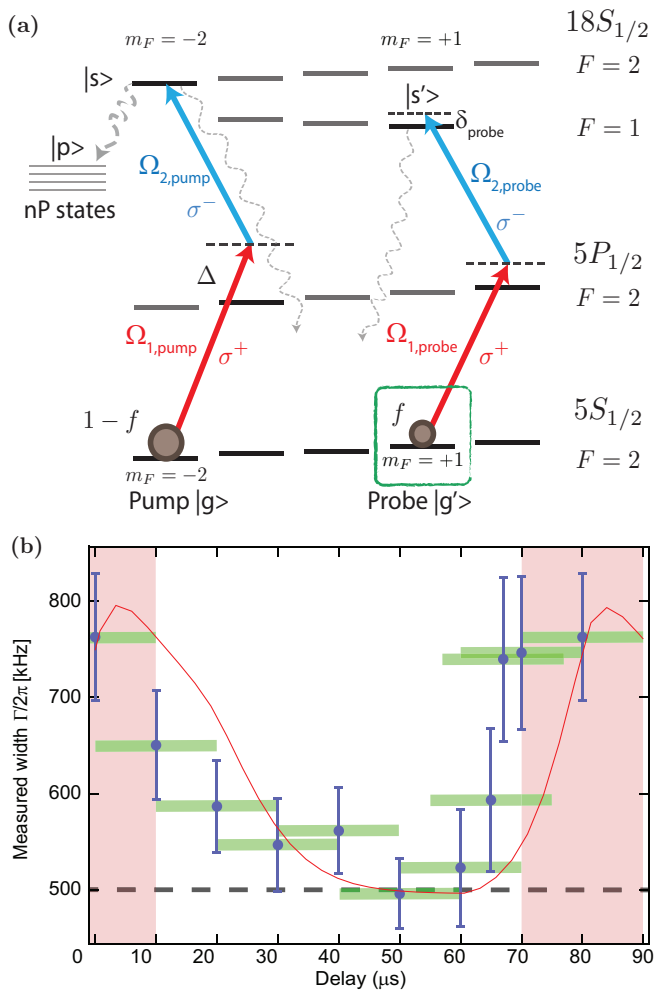}
\caption{\textbf{The $\mathbf{\Delta m_F = 3}$ case. (a)} Pumping scheme for two states separated by $\Delta m_F = 3$. The pump population is driven resonantly to $\ket{s}=\ket{18S, 2,-2}$, while the probe transition is driven to $\ket{s^\prime}=\ket{18S,1,+1}$. 
\textbf{(b)} Time dependence of the pump-probe experiment. The red zones represent the times the pump is on, while the green bands show the probe pulse width. Again we observe a spectral width increase when in-phase, relative to out-of-phase. This plot spans the full period $T$ and shows an asymmetry between having the probe after ($0-40~\si{\micro\second}$) or before ($40-80~\si{\micro\second}$) the pump. A rate equations theory (see Eqs.~(\ref{Rate1}-\ref{Rate-R})) shows (red line) the correct shape, including asymmetry, but consistently has a delay of $\sim \si{10~\micro\second}$.}
\label{result_DmF3}
\end{figure}

To study the dynamics of contaminant p-state population, we measured the probe spectral width as a function of pump-probe delay $\Delta t$ over a full period $T$. For this experiment, we chose $\ket{s^\prime}=\ket{18S, F=1, m_F=1}$ as a probe state. The $\Delta m_F = 3$ between the probe and the pump s-states implies a process of the type $\ket{p,s'}\leftrightarrow\ket{s'',p'}$ where $|m_{F,p}-m_{F,s''}|\leq 1$ and $|m_{F,s'}-m_{F,p'}|\leq 1$, and where $\ket{p}\neq\ket{p'}$ and $\ket{s'}\neq\ket{s''}$( since $|m_{F,p}-m_{F,s'}|\geq 2$). In order to get a significant two-photon Rabi frequency for both pump and probe transitions we used as before an intermediate state detuning $\Delta\approx 2\pi\times\si{240\mega\hertz}$, but this time blue of the $\ket{5S, F=2}\rightarrow \ket{5P_{1/2}, F=2}$ transition (Fig.~\ref{result_DmF3}(a)). Here $t_p=\si{20~\micro\second}$ and $T=\si{80~\micro\second}$. Fig.~\ref{result_DmF3}(b) shows the width as a function of the delay. The largest value is observed for overlapping pulses, here too indicating that the pump pollutant population dephases the probe population. We note that the strength of the dipole exchange is almost the same as for the $\ket{18S, F=1, m_F=-1}$ probe state, despite the fact that $\Delta m_F = 3$. 

The shape and amplitude of the time-dependent signal are captured by a nonlinear rate-equation model, described below. It uses a single fit parameter, the effective interaction strength $C_3^\textrm{cross}$ between the pump p-state (s-state) and the probe s-state (p-state). Interestingly, the model is consistently $\sim \si{10~\micro\second}$ delayed with respect to the experiment; see the red solid lines in Fig.~\ref{result_pulsed_delay}(b) and in Fig.~\ref{result_DmF3}(b). As discussed below, this is likely a failure of the model's homogeneous mean-field assumptions to capture early time dynamics due to strong local correlations (cluster dynamics) from which the pollutant population is seeded.

The cluster dynamics suggests a finite growth time. Therefore, for an excitation pulse short enough to avoid the p-state creation, the self-broadening might be avoided. Using a single $\ket{5S \rightarrow 18S, F=2, m_F=-2}$ transition, we measured the linewidth as a function of pulse width $t_p$, using pulse trains in a similar fashion as shown in Fig.~\ref{Schematics}(b) but without probe light. Spectra were taken at different $t_p$ while keeping the total excitation time $t_p N_p$ constant (where $N_p$ is the number of pulses). To allow all excited population to decay in between pulses, the dark time was kept long compared to $\tau_0$: $t_\textrm{dark} \simeq 20\tau_0 \simeq 10\Gamma_\textrm{nP}^{-1}$. The observed width is presented in Fig.~\ref{result_NISTpulses}. There is a clear decrease of the width at short times, up to 3 times narrower than the steady-state observations. At very short pulse widths, however, Fourier broadening begins to dominate, and the narrowing is limited to pulse times $t_p\gtrsim\si{2~\micro\second}$. We observe a time of about $\si{10~\micro\second}$ to reach steady-state and infer a timescale of a few hundreds of nanoseconds for the onset of the broadening. The nonlinear rate-equation model (described below) captures the steady-state dephasing (outside the plot range) and the global shape of the time evolution. The speed at which steady-state is approached is however consistently underestimated by the model. As discussed below, this slow rise time is likely the result of the homogeneous mean-field assumption used here.

\subsection{Nonlinear rate equation model}

We compare the self-broadening data to a simple homogeneous non-linear rate-equation model, which is a good approximation for situations with strong decoherence \cite{ates2007many,honing2013steady}. We approximate the system with three coupled populations (ground (g), Rydberg (18S), pollutant (nP)), treating all the pollutant states as a single effective state:
\begin{align}
&\dot{N}_g=(N_\textrm{18S}-N_g)R + \Gamma_0 b_1 N_\textrm{18S} + \Gamma_\textrm{nP} b_3 N_\textrm{nP}, \\ \label{Rate1}
&\dot{N}_\textrm{18S} = (N_g-N_\textrm{18S})R - \Gamma_0 N_\textrm{18S}, \\
&\dot{N}_\textrm{nP} = b_2 \Gamma_0 N_\textrm{18S} - \Gamma_\textrm{nP} N_\textrm{nP},
\end{align}
where $b_1=0.49$ is the branching ratio from the $18S$ states back to $\ket{g}=\ket{5S, F=2, m_F=-2}$, $b_2=\sum_\textrm{nP} b^\textrm{nP}_2 = 0.18$ is the branching ratio of the decay from $\ket{18S, F=2, m_F=-2}$ to the effective pollutant state, and $b_3=0.55$ is the branching ratio from the effective pollutant state back to $\ket{g}$. Decay of the total population represents atom loss and optical pumping to detection ground states. The interactions are modeled under a homogeneous mean-field assumption that the $18S$ dephasing rate depends on the typical density of pollutant atoms
\begin{equation}
\Gamma(t) = \Gamma_0 + C_3 \rho_0 N_\textrm{nP}(t), \label{Rate-Gamma}
\end{equation}
where $\rho_0$ is the density of atoms (in any state), $N_\textrm{nP}$ is the fraction of these atoms in the pollutant state, and $C_3\simeq 2\pi\times 35~\si{\mega\hertz~\micro\meter^3}$ is the average of the $C_3^\textrm{(nP)}$ interaction strengths used in Eq.~(\ref{Eq-AB}). The dephased excitation rate, including broadening and saturation, is then taken to be the time-dependent form
\begin{equation}
R(t) = \frac{\Gamma(t)}{2}\frac{2\Omega^2}{4\delta^2 + \Gamma^2(t)}. \label{Rate-R}
\end{equation}
This model is consistent with the steady-state limit presented in Eq.~\ref{Eq-AB}. As shown in the Appendix, the model used for the cross-broadening experiment consists of a pair of such 3-level systems coupled to each other by adding a cross-interaction term $C_3^\textrm{cross} \rho_0 N_\textrm{nP}^\textrm{pump (probe)}$ in the systems' respective dephasing rates $\Gamma^\textrm{probe (pump)}$. 

To compare with the experiment, we calculated (both for the self- and cross-broadening cases) the time dependence of the populations, starting with all atoms in the ground state(s). The optically pumped population that serves as our experimental signal is determined from the number of atoms remaining in the probe ground state at the end of the pulses, $N_g(t)$. This is repeated at a range of detunings to construct a simulated spectrum, from which we determine a width by fitting a Lorentzian.

\begin{figure}[!t]
\includegraphics[width=8.6cm]{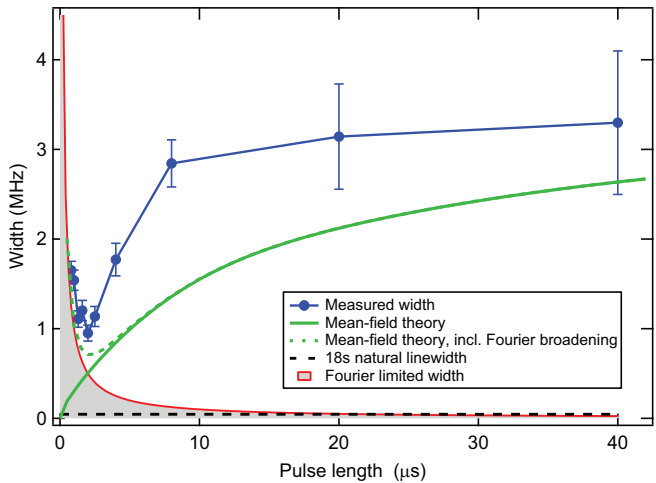}
\caption{\textbf{Stroboscopic excitation on a single transition.} Observed self-broadened width (blue) of the pump transition as a function of the individual pulse width. Here a single $\ket{18S,2,-2}$ population is excited with a Rabi frequency $\Omega=2\pi\times\si{66\kilo\hertz}$. Below $t_p \approx 3\tau_0$ (where $\tau_0$ is the $18S$ lifetime), the broadening decreases, with a reduction of the width by a factor of $\sim 3$. However, the natural linewidth (black dashed line) cannot be recovered via this scheme due to the fundamental Fourier limit (red line, delimiting the forbidden region in grey). Here too the homogeneous mean-field prediction (green line) is slower than observed. The dashed green line shows the prediction for homogeneous mean-field plus Fourier broadening.}
\label{result_NISTpulses}
\end{figure}

For the delayed-pulse cross-broadening experiment, the model's results are shown as a red line in Fig.~\ref{result_pulsed_delay}(b) and in Fig.~\ref{result_DmF3}(b). It correctly predicts the overall time dependence with only one free parameter, $C_3^\textrm{cross}$, for which the extracted value is $C_3^\textrm{cross} \simeq C_3 / 10 = 2\pi\times 3.5~\si{\mega\hertz~\micro\meter^3}$. It also captures the observed asymmetry on turn-on and turn-off of the pump in Fig.~\ref{result_DmF3}(b), but is delayed relative to the data by $\sim \si{10~\micro\second}$. We note that the time scale for relaxation to the probe-only width takes somewhat longer than the $nP$ lifetime, which is indicative of the dynamics of facilitated excitation~\cite{letscher2016bistability}.

To gain better insight into this behavior, we look at the model's result for a pump-only pulsed excitation and compare it to the equivalent self-broadening experiment. This is shown as the green line in Fig.~\ref{result_NISTpulses}. Although the model tends to capture the late-times widths~\cite{AB, jeremy} (the experimental value is recovered at long times, outside of the plot range in Fig.~\ref{result_NISTpulses}), it consistently gives slower dynamics than observed experimentally. \ZZ The experimental timescales are about five times faster than the model. ZZ Since the growth of the pollutant population is expected to occur through an aggregation process similar to~\cite{Urvoy2015}, strong correlations arise quickly after the first pollutant atom is created. A single atom in an $nP$ state can shift some fraction of surrounding atoms into resonance, leading to fast excitation in an anti-blockade, clustered dynamic \cite{letscher2016bistability,letscher2017anomalous}. Therefore, as recently highlighted~\cite{jeremy}, inhomogeneity and high-order correlations are important.

\begin{figure}[!t]
\includegraphics[width=8.6cm]{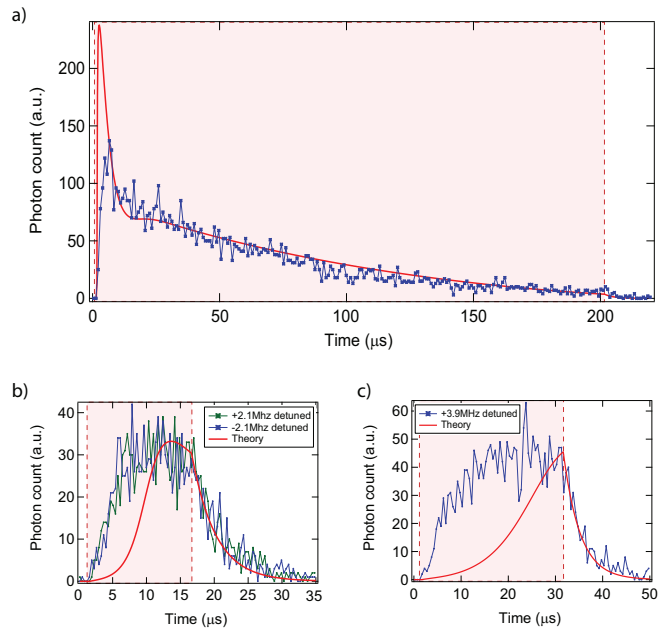}
\caption{\textbf{Photon counting on a single transition.} Observed $5P_{3/2}$-$5S$ fluorescence (blue trace) versus time. The detected fluorescence is proportional to the $18S$ population. Here a single $\ket{18S,2,-2}$ population is excited with a Rabi frequency $\Omega=2\pi\times\si{140\kilo\hertz}$ (light red areas). The red line is the the model. \textbf{(a)} Resonant case: an overshoot is visible in both the data and the model, signaling the presence of the $nP$ population. The model being slower to produce pollutants, the overshoot in the model has more time to grow at the full single-particle rate than in reality. \textbf{(b),(c)} Detuned case: since the facilitated excitation relies on a pollutant population, the $18S$ population is slower to rise in the model than experimentally observed, confirming the faster-than-homogeneous-mean-field property of the avalanche dephasing. The experimentally observed broadening being symmetric, the sign of the detuning does not change these observations.}
\label{photons}
\end{figure}

This is confirmed by comparing the calculated $18S$ population dynamics to the corresponding experimental observation, obtained with fluorescence photon counting. The detected fluorescence (proportional to $18S$ population) vs. time is presented in Fig.~\ref{photons} for detunings $\delta/2\pi = 0~\si{\mega\hertz}, \pm 2.1~\si{\mega\hertz}$, and $3.9~\si{\mega\hertz}$ on a single transition (pump-only). Here we chose $\Omega = 2\pi\times 140~\si{\kilo\hertz}$. In all the data, the dephasing's effect appears earlier than in the homogeneous mean field theory. For detuned light (Fig.~\ref{photons}(b) and (c)), the calculated facilitated excitation in the model happens on a timescale at least twice as long as in the experiment. For the resonant case (Fig.~\ref{photons}(a)), the decrease of the pumping rate (blockade effect) also happens later in the model, which manifests itself as a large overshoot at early times. Such overshoot is also present in the data, but with a lower amplitude, indicating that the correlated dynamics starts the dephasing earlier. It is important to note that the presence of an overshoot in the data implies a finite time for the dephasing to occur. This demonstrates the existence of a third state (or set of states), since $18S$-$18S$ interactions cannot produce such an overshoot. This state, the existence of which is directly demonstrated here, is necessarily the $nP$ population, as verified by the experimental observations~\cite{gallagher2005,AB,Killian16,Bloch} in which it was originally inferred.

\section{Approaches to mitigate broadening}
Given the range of applications of Rydberg atoms, it is worthwhile to consider possible approaches to minimize the impact of this decoherence mechanism.

\subsection{Stroboscopic excitation}
\label{sec:Strobo}
The results of Fig.~\ref{result_NISTpulses} and Fig.~\ref{photons} indicate that the contaminant dephasing can be partially avoided by pulsing the Rydberg coupling on a short enough timescale $t_p$, followed by sufficiently long dark time $t_\textrm{dark}$. To allow for decay of any detrimental population, $t_\textrm{dark}=A \tau_0$, with $A\gg 1$ (to avoid the runaway process on the next pulse). The average time $\tau_c$ until the first contaminant-state atom appears provides an estimate of the actual coherent time available before the avalanche excitation is triggered. The associated rate $\tau_c^{-1}$ is roughly the number of atoms in the Rydberg state, $N\left(\frac{\Omega}{2\delta}\right)^2$, times the rate at which this state decays to contaminant states. Therefore, given a sample with a total of $N$ atoms participating in the dressed excitation, an estimate for $\tau_c$ is~\cite{AB}:
\begin{equation}
\frac{\tau_c}{\tau_0} = \frac{4\delta^2}{\Omega^2}\frac{1}{b_\textrm{nL}N}\equiv\frac{N_c}{N},
\end{equation}
\noindent
where $b_\textrm{nL}$ is the sum of the branching ratios from the general $nL$ Rydberg-dressed state (with $n$ the principal quantum number and $L=0,1,2,$... the angular momentum) to the contaminant states contributing to the effective interaction volume $\beta$. $N_c = 4 \delta^2/ \Omega^2 b_\textrm{nL}$ is the number of atoms above which stroboscopic approaches will start to significantly diminish the dressing potential strength. The effective $N$ may be suppressed (at short times) for systems in which a significant fraction of atoms are within a $\ket{s}$-$\ket{s}$ van der Waals blockade distance. However the $\ket{s}$-$\ket{p}$ dipole-dipole interaction is much stronger and longer ranged than the van der Waals interaction, and once the contaminant $\ket{p}$ states are produced, the dipole broadening mechanism dominates. We note that since $\tau_0$ increases with principle quantum number $n$ (and $b_\textrm{nL}$ increases weakly with $n$), increasing $n$ will improve the overall time scale~\cite{saffman} before decoherence starts. Nevertheless, the decrease of $\tau_c$ with increasing atom number $N$ is still present even at large $n$. Additionally, the situation is reversed after the first pollutant has appeared: $C_3$ increases with $n$ and therefore once started the dephasing is stronger at higher $n$.
 
For experiments where the observable does not rely on correlations being preserved during the dark time and is on a much longer timescale than the total stroboscopic period $T$ (which will not always be the case), the dressed interaction during $t_p<\tau_c$ could be averaged across many pulses. In principle it would then be possible to avoid the avalanche dephasing. The available average interaction for a given coupling $\Omega$ and detuning $\delta$ would be decreased, however, by the duty cycle $t_p/T < \tau_c/(\tau_c + A \tau_0)$. The short-distance dressed interaction energy $U =\frac{\Omega^4}{8 \delta^3}$~\cite{honer2010collective} is then limited to
\begin{equation}
U^* \lesssim \left(\frac{\Omega^4}{8 \delta^3}\right)  \frac{N_c/N }{A + N_c/N }.
\end{equation}
For $N_c/N \gg A$, the stroboscopic approach gives a time-averaged interaction energy only slightly suppressed, $U^* \lesssim U$. In the opposite limit $N_c/N \ll A$ (which includes $N\gtrsim N_c$ since $A\gg 1$), the average interaction scales as $U^* \lesssim U N_c/A N  \ll U$.  For experimentally reasonable parameters of $\delta/\Omega \simeq 10$ and $b_\textrm{nL} \simeq 0.2$, $N_c \simeq 2000 $.  Fourier broadening provides a further fundamental limit to the stroboscopic approach. Avoiding Fourier-broadened excitation requires $\delta^{-1} \ll t_p < \tau_c $, which sets the additional constraint $N\ll N_c \delta/\Gamma_0$. 

\subsection{Cryogenic temperatures}
\begin{figure}[]
  \includegraphics[width=8.6cm]{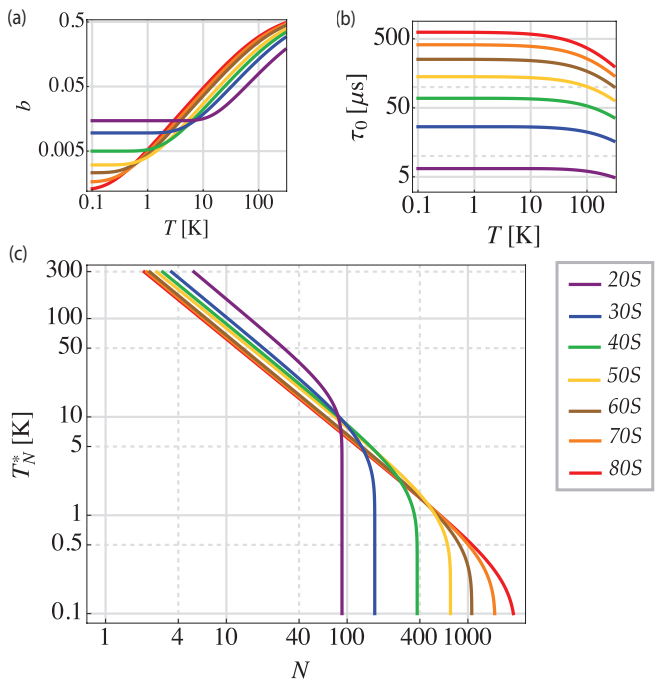}
\caption{\textbf{Temperature scaling. (a)} Temperature dependence of the sum of branching ratios $b$ from the different $nS$ states to the contaminant $nP$ states playing a significant role in the avalanche dephasing process. \textbf{(b)} Effective lifetime $\tau_0$ (determined by spontaneous emission and blackbody radiation-induced transitions) for different $nS$ $^{87}$Rb Rydberg states as a function of the ambient temperature. (c)  Temperature $T^*_N$ needed to compensate for the dephasing effect as a function of $N$. The sharp drop stems from the intrinsic  branching ratio to contaminant states: even at zero temperature, there is a non-zero chance to produce a contaminant atom.}
\label{fig:cthcal_4}
\end{figure}

The time $\tau_c$ before the first contaminant-state atom appears depends on the ambient temperature through $b_\textrm{nL}$ and $\tau_0$, as shown in Fig.~\ref{fig:cthcal_4}(a) and Fig.~\ref{fig:cthcal_4}(b) for example $^{87}$Rb $nS$ Rydberg states. These plots were obtained using the estimates from Ref.~\cite{Beterov2009} and the quasiclassical formulas in Ref.~\cite{Dyachkov1994}. It is clear that lowering the ambient temperature increases $\tau_c$, as well as increases the value of $N_c$ for the stroboscopic approach. This is mainly due to the reduction of the branching ratio to nearby states $b_{nL}$. At room temperature, there is significant blackbody-driven decay to nearby states rather than by radiative decays to low-$n$ states, due to the small energy difference between neighboring Rydberg states. Suppressing blackbody radiation at low temperatures makes the spontaneous decay to low-lying states the dominant channel.

Avalanche dephasing limits the coherence below the single-particle scattering time when $\tau_c < \frac{4\delta^2}{\Omega^2}\tau_0$. For a given atom number $N$, we estimate the temperature $T_N^*$ at which the dephasing time $\tau_c$ becomes smaller than the room temperature single-particle scattering time, namely the temperature at which $\tau_c (T_N^*)= \frac{4\delta^2}{\Omega^2}\tau_0(300K)$. $T_N^*$ vs. $N$ is shown in Fig.~\ref{fig:cthcal_4}(c).

Rydberg-dressing proposals~\cite{pupillo2010,Glaetzle2012,Bouchoule2002,Glaetzle2014} and experiments~\cite{Bloch} making use of samples with $N\sim 30-100$ would need an ambient temperature near $\si{10~\kelvin}$ to make up for the avalanche dephasing effect. The detrimental effects of contaminant states on systems with even lower atom number ($N\sim 10$, see Refs.~\cite{vanBijnen2015,Lee2013a}), already small at room temperature, would become negligible for temperatures around liquid nitrogen, $77~\si{\kelvin}$. On the other hand, avoiding contaminant dephasing in proposals relying on $N \gtrsim 10^3$, as in Ref.~\cite{AB}, does not seem feasible by going to lower temperatures. This is visible in Fig.~\ref{fig:cthcal_4}(c) as a sharp drop to $T=\si{0~\kelvin}$ for the temperature needed to avoid the avalanche dephasing at high atom numbers, due to the always-finite value of spontaneous radiative decay rates.

A possible solution would be to suppress spontaneous decay with an off-resonant cavity in a sufficiently cold cryogenic environment (for example, a dilution fridge techniques can reach $T\leq 0.1\si{\kelvin}$), eg. with a superconducting cavity. For a conductor separation smaller than the spontaneous emission wavelength, the Rydberg excitation cannot decay to vacuum modes and spontaneous decay is suppressed~\cite{kleppner1981inhibited,hulet1985inhibited}. Suppressing both blackbody and spontaneous decay channels has the capacity to avoid the avalanche dephasing issue, despite substantial technical challenges.

\subsection{Other approaches}

\textbf{Quenching contaminant atoms} - One possibility to limit the impact of pollutants is to significantly shorten their lifetime so they do not live long enough to begin the avalanche process. This could be achieved, for example, by laser coupling the contaminant states to lower-lying, short-lived states. This approach should work in principle, but is challenging to realize experimentally because of the large number of states with appreciable $C_3$ coefficient with the targeted dressing state, requiring quenching of many states at once. Indeed, for a $nS$ initial Rydberg state, there are at least four significant pollutant states: $nP_{1/2}$, $nP_{3/2}$, $(n-1)P_{1/2}$ and $(n-1)P_{3/2}$, plus all their hyperfine levels (if they are spectrally resolved). For an initial Rydberg state with $L\geq 1$, the number of significantly contributing neighbor states increases rapidly, requiring a large number of quenching lasers.

\textbf{EIT dressing schemes} - The idea of taking a new approach to Rydberg dressing~\cite{pohl} has also been discussed as a potential way out of the avalanche dephasing~\cite{saffman}. The idea presented in Ref.~\cite{pohl} uses resonant light in a two-atom EIT condition to achieve a Rydberg dressing-like interaction potential. The double excitation to a Rydberg level of the atom pair can be tuned to a near-dark state (EIT condition). This reduces the Rydberg fraction of the final state, while achieving the same hallmarks as traditional Rydberg dressing. This method shows an increase of about an order of magnitude in the interaction strength compared to the usual dressing scheme for strontium, and is about as effective as usual dressing for rubidium. 

This scheme suffers as well from the avalanche dephasing: the dressing interaction stems from a non-zero Rydberg fraction, which can trigger the avalanche process in large ensembles. The data shows clear signs of early decoherence~\cite{pohl,aman2016trap}. The tenfold improvement in $U$ for strontium does however mean that fewer atoms are needed to achieve the same interaction energy as with the classic scheme. This presents an additional way to reduce somewhat the impact of the avalanche decoherence, at least for Rydberg systems with narrow intermediate state transitions.

\textbf{Post-selection} - Once triggered, the avalanche process occurs very quickly, resulting in dramatic ground state excitation and loss. The authors in Ref.~\cite{Bloch} took advantage of this fact by post-selecting the experimental runs where the avalanche never occurred. While this approach does not effectively increase the average time scale for the experiment or change its scaling with $N$, it does allow for selecting contaminant-free data, at the expense of a decreased data rate. In Ref.~\cite{Bloch}, even for a relatively short experimental time $t \sim \tau_0\left(n\right)$ (the state lifetime) and for $N\simeq 100$, half the experimental runs were corrupted. We note that post-selection could increase the parameter regime where stroboscopic approaches work, relaxing somewhat the constraints on that approach.

\section{Conclusion}
In conclusion, we presented a set of experiments studying the rapid onset of decoherence systematically observed in large Rydberg ensembles. The data supports the proposed explanation~\cite{AB} that runaway production of atoms in contaminant Rydberg states, triggered by blackbody decay, is the source of the broadening. This decoherence mechanism is important to consider, as it should be present in any experiment with steady-state Rydberg population, such as off-resonantly dressed Rydberg  systems or  Rydberg-based quantum information processing. Our results also point towards the difficulty of avoiding the issue. We provide and discuss several ideas to reduce the impact of this effect, among which cryogenic operation has the highest figure of merit, especially if used in conjunction with cavity techniques, despite the inherent technical challenges. None of the proposed solutions, however, offer an easy way to evade the avalanche decoherence, but a combination of approaches may provide a way to mitigate the effect on current experiments. 

On the other hand, this mechanism can be used to explore highly-correlated, many-body dynamics~\cite{jeremy}, such as anti-blockade on both sides of the Rydberg transition (``anomalous facilitation'' of excitation~\cite{letscher2017anomalous}), and related cluster growth dynamics~\cite{letscher2016bistability,petrosyan2013spatial}. The large (contaminant-induced) anomalous facilitation and the faster-than-homogeneous-meanfield excitation rate seen here provide strong indication of cluster growth dynamics in our system. In addition, such interaction-induced dephasing may have practical applications in few Rydberg systems, such as in the isolation of single-Rydberg excitations for entanglement generation~\cite{bariani2012dephasing}.

\begin{acknowledgments}
This work was partially supported by ARL-CDQI, AFOSR, ARO MURI, NSF QIS, ARO, and NSF PFC at JQI. T.B. acknowledges the support of the European Marie Sk\l{}odowska-Curie Actions (H2020-MSCA-IF-2015 Grant 701034).
\end{acknowledgments}

\section{Appendix}

\subsection{Cross-broadening rate equation model}

As described in the main text, the simple three-level rate-equations model was extended to six states in order to account for the two (pump and probe) species and the cross-interaction. This more general model takes into account both the effect of the pump on the probe and the effect of the probe on the pump. The full set of equations is:

\begin{widetext}
\begin{align}
&\dot{N}_\textrm{1,g}(t)=(N_\textrm{1,18S}(t)-N_\textrm{1,g}(t))R_1(t)+b_\textrm{1,1} \Gamma_0 N_\textrm{1,18S}(t)+ b_\textrm{3,1} \Gamma_\textrm{nP} N_\textrm{1,nP}(t)-\Gamma_{1,D} N_\textrm{1,g}(t), \\
&\dot{N}_\textrm{1,18S}(t) = (N_\textrm{1,g}(t)-N_\textrm{1,18S}(t))R_1(t) - \Gamma_0 N_\textrm{1,18S}(t), \\
&\dot{N}_\textrm{1,nP}(t) = b_\textrm{2,1} \Gamma_0 N_\textrm{1,18S}(t) - \Gamma_\textrm{nP} N_\textrm{1,nP}(t),\\
&\dot{N}_\textrm{2,g}(t)=(N_\textrm{2,18S}(t)-N_\textrm{2,g}(t))R_2(t)+ b_\textrm{1,2} \Gamma_0 N_\textrm{2,18S}(t)+b_\textrm{3,2} \Gamma_\textrm{nP} N_\textrm{2,nP}(t)-\Gamma_{2,D} N_\textrm{2,g}(t), \\
&\dot{N}_\textrm{2,18S}(t) = (N_\textrm{2,g}(t)-N_\textrm{2,18S}(t))R_2(t) - \Gamma_0 N_\textrm{2,18S}(t), \\
&\dot{N}_\textrm{2,nP}(t) = b_\textrm{2,2} \Gamma_0 N_\textrm{2,18S}(t) - \Gamma_\textrm{nP} N_\textrm{2,nP}(t),
\end{align}
\end{widetext}
with the same mean-field assumption that the $18S$ dephasing rate depends on the typical density of pollutant atoms:
\begin{align}
&R_1(t) = \frac{\Gamma_1}{2} \frac{2\Omega^2}{\Gamma_1^2+4\delta_1^2}, \\
&R_2(t) = \frac{\Gamma_2}{2} \frac{2\Omega^2}{\Gamma_2^2+4\delta_2^2}, \\
&\Gamma_1 = \Gamma_0 + C_3 \rho_0 N_\textrm{1,nP} + C_3^\textrm{Cross} \rho_0 N_\textrm{2,nP}, \\
&\Gamma_2 = \Gamma_0 + C_3 \rho_0 N_\textrm{2,nP} + C_3^\textrm{Cross} \rho_0 N_\textrm{1,nP},
\end{align}
where $b_\textrm{1,i}$ is the branching ratio back to the ground state of specie $i$, $b_\textrm{2,i}=\sum_\textrm{nP} b^\textrm{nP}_\textrm{2,i}$ is the branching ratio of the decay from Rydberg state to the effective pollutant state, here taken as equal for $i=1$ (pump) and $i=2$ (probe),  and $b_\textrm{3,i}$ is the branching ratio from the effective pollutant state of specie $i$ back to the ground state of the same specie. $\delta_i$ is the 2-photon detuning. Since the pump is always resonant in the experiment, we used $\delta_1 = 0~\si{\mega\hertz}$. The terms $-\Gamma_{i,D} N_\textrm{i,g}$ in the two ground state equations correspond to the off-resonant scattering from the $5P_{1/2}$ state, with scattering rate $\Gamma_{i,D} = \left(\frac{\Omega_{1,i}}{2\Delta_i}\right)^2 \Gamma_{5P}$ (with $\Gamma_{5P}=2\pi\times 6~\si{\mega\hertz}$). Here $C_3 \simeq 2\pi\times 34~\si{\mega\hertz~\micro\meter^3}$ and $C_3^\textrm{cross}$ is a fit parameter, found to be about $2\pi\times\si{3~\mega\hertz}$. $\Gamma_\textrm{nP}$, $C_3$ and $C_3^\textrm{cross}$ are taken to be the same for populations 1 and 2.

%
%
%

%

\end{document}